\documentclass[journal]{utilities/IEEEtran}

\ifCLASSINFOpdf
\else
   \usepackage[dvips]{graphicx}
\fi
\usepackage{url}

\usepackage[utf8]{inputenc}
\usepackage{graphicx}
\usepackage{amsmath,graphicx}
\usepackage{amssymb}

\usepackage{pgfplots}
\DeclareUnicodeCharacter{2212}{−}
\usepgfplotslibrary{groupplots,dateplot}
\usetikzlibrary{patterns,shapes.arrows}
\pgfplotsset{compat=newest}
\usepackage{mathtools}
\usepackage{lipsum}
\usepackage{balance}
\newcommand{\fref}[1]{Fig.~\ref{#1}}

\newcommand{\eref}[1]{(\ref{#1})}
\newcommand{\sref}[1]{Section~\ref{#1}}

\usepackage{utilities/notations}
\usepackage[acronym]{glossaries}
\usepackage{bm}
\newacronym{asr}{ASR}{automatic speech recognizer}
\newacronym{brir}{BRIR}{binaural RIR}
\newacronym{bsim}{BSIM}{binaural speech intelligibility measure}
\newacronym{bstoi}{BSTOI}{binaural STOI}
\newacronym{cpc}{CPC}{contrastive predictive coding}
\newacronym{dbstoi}{DBSTOI}{deterministic BSTOI}
\newacronym{ec}{EC}{equalization-cancellation}
\newacronym{ema}{EMA}{exponential moving averages}
\newacronym{fft}{FFT}{fast Fourier transform}
\newacronym{gru}{GRU}{gated recurrent neural network}
\newacronym{lcc}{LCC}{Pearson's correlation coefficient}
\newacronym{mel}{MelSpec}{mel-spectrogram}
\newacronym{mlp}{MLP}{multi-layer perceptron}
\newacronym{mse}{MSE}{mean-squared error}
\newacronym{mtd}{MTD}{mean temporal distance}
\newacronym{relu}{ReLU}{rectified linear unit}
\newacronym{rir}{RIR}{room impulse response}
\newacronym{rnn}{RNN}{recurrent neural network}
\newacronym{seqpool}{SeqPool}{sequence pooling}
\newacronym{srcc}{SRCC}{Spearman's rank correlation coefficient}
\newacronym{si}{SI}{speech intelligibility}
\newacronym{sii}{SII}{speech intelligibility index}
\newacronym{sip}{SIP}{speech intelligibility prediction}
\newacronym{snr}{SNR}{signal-to-noise ratio}
\newacronym{srt}{SRT}{speech reception threshold}
\newacronym{sti}{STI}{speech transmission index}
\newacronym{stoi}{STOI}{short-time objective intelligibility}
\newacronym{vq}{VQ}{vector quantization}
\newacronym{vqcpc}{VQ-CPC}{vector quantized contrastive predictive coding}

\begin{document}

\title{Non-Intrusive Binaural Speech Intelligibility Prediction From Discrete Latent Representations}

\author{Alex F. McKinney~\IEEEmembership{Student~Member,~IEEE} and Benjamin Cauchi~\IEEEmembership{Member,~IEEE}
\thanks{
This work was conducted as part of the RISE exchange program funded by the German Academic Exchange Service (DAAD) and supported by the project Augmented Auditory Intelligence (A2I) funded by the German Ministry of Education and Research (BMBF) under grant number 16SV8594.}
\thanks{Alex F. McKinney is with the Department of Computer Science of Durham
University, United Kingdom (e-mail: alexander.f.mckinney@durham.ac.uk).}
\thanks{Benjamin Cauchi is with the OFFIS e.V. Institute for Information Technology,
Oldenburg, Germany (e-mail: benjamin.cauchi@offis.de).}}

\maketitle

\glsunset{vqcpc}
\glsunset{stoi}
\glsunset{mse}
\begin{abstract}
Non-intrusive \gls{si} prediction from binaural signals is useful in many
applications.  However, most existing signal-based measures are designed to be
applied to single-channel signals.  Measures specifically designed to take into
account the binaural properties of the signal are often intrusive --
characterised by requiring access to a clean speech signal -- and typically rely
on combining both channels into a single-channel signal before making
predictions. This paper proposes a non-intrusive \gls{si} measure that computes
features from a binaural input signal using a combination of \gls{vq} and
\gls{cpc} methods.  \gls{vqcpc} feature extraction does not rely on any model of
the auditory system and is instead trained to maximise the mutual information
between the input signal and output features. The computed \gls{vqcpc} features
are input to a predicting function parameterized by a neural network. Two
predicting functions are considered in this paper.  Both feature extractor and
predicting functions are trained on simulated binaural signals with isotropic
noise. They are tested on simulated signals with isotropic and real noise. For
all signals, the ground truth scores are the (intrusive) deterministic binaural
\gls{stoi}.  Results are presented in terms of correlations and \gls{mse} and
demonstrate that \gls{vqcpc} features are able to capture information relevant
to modelling \gls{si} and outperform all the considered benchmarks -- even when
evaluating on data comprising of different noise field types.




\end{abstract}
\glsresetall  


\IEEEpeerreviewmaketitle

\section{Introduction}
\label{sec:intro}
\glsunset{vqcpc}

\Gls{si} prediction aims to predict the ability of an average listener to
comprehend speech within a signal -- potentially corrupted by noise, reverberation
or processing artefacts. \Gls{si} is defined as the number of words or phonemes
that can be correctly identified by assessors. It is often reported using the
\gls{srt} defined to be the level of degradation for which only 50\% of the
speech items are correctly identified~\cite{doire2017b}.
Listening tests are typically considered the gold standard to measure \gls{si}.
However, these tests are costly, time-consuming and cannot be applied in
real-time, making the use of signal-based measures necessary. Signal-based
measures of speech intelligibility can be categorized as either intrusive or
non-intrusive. The computation of intrusive measures requires a clean
reference signal in addition to the test signal, whereas non-intrusive measures
can be computed from the test signal only.  Consequently, only non-intrusive
measures can be applied in real-time settings. Additionally, \gls{si} largely
depends on the presence of binaural cues~\cite{bronkhorst2000} and so \gls{si}
measures should be developed to incorporate them.

Most signal-based measures of \gls{si} were originally designed for telephony
applications and applied only to single-channel signals.  Among these measures
the articulation index~\cite{French1947}, the \gls{sti}~\cite{steeneken1980},
the \gls{sii}~\cite{ANSI_S3_5_1997}, the \gls{stoi}~\cite{taal2011} and
mutual-information-based techniques, such as the algorithm proposed
in~\cite{taghia2014}, are intrusive.  Non-intrusive measures include a
non-intrusive extension of the \gls{stoi}~\cite{andersen2017,sorensen2018}, that
relies on estimating the amplitude envelope of the clean speech from the input
signal, and measures relying on machine learning techniques. Some measures use a
trained speech recognizer as proposed in~\cite{spille2018,schaedler2020} or a
neural network trained to predict \gls{si} from a sequence of spectral
features~\cite{zezario2020}.

The aforementioned measures use a single-channel signal as input. Certain
measures instead make predictions from binaural signals, aiming to take
into account the relevance of binaural cues in modelling the intelligibility of
speech signals.
Notably, several binaural measures rely on simple \gls{ec} models~\cite{durlach1963},
often being combined with the \gls{sii}~\cite{beutelmann2010,lavandier2010}.
However, these measures do not take into account the impact that non-linear
processing has on \gls{si}. The \gls{bstoi} uses an \gls{ec} model to combine
both channels of the binaural signal into a single-channel signal used as input
to the \gls{stoi} measure. \Gls{bstoi} was later refined into the \gls{dbstoi}
that produces a deterministic output~\cite{andersen2018a}. Both \gls{bstoi} and
\gls{dbstoi} are intrusive. A non-intrusive measure has been proposed
in~\cite{rossbach2021}, where the blind binaural preprocessing stage
from~\cite{hauth2020} is used to process the binaural signal into a
single-channel signal that is then input to an \gls{asr}. The \gls{si} is finally
predicted by applying a trained mapping between the \gls{mtd} -- a
representation of the \gls{asr} error~\cite{hermansky2013} -- and the \gls{srt}.

Some measures do not rely on on any model of the auditory system and input features
to a predicting function that needs to be trained. Such methods include the use of both short- and long-term features input to a classification and regression tree~\cite{sharma2016} or the use of \gls{stoi} like features as input to a convolutional neural network~\cite{andersen2018b}.
The measure proposed in this paper applies a similar approach but uses features that
are computed as a latent representation of the input binaural
signal using a combination of \gls{cpc}~\cite{oord2019} and
\gls{vq}~\cite{vanniekerk2020} methods. Previous uses of \gls{cpc} in audio
applications were limited to single-channel audio
inputs~\cite{oord2019,schneider2019,baevski2020} and images~\cite{oord2019}.
\gls{cpc} models excel at representing ``slow features'' that span many time
steps~\cite{oord2019} which we believed would make it suited for \gls{sip}. This is
in contrast to other self-supervised methods such as autoencoders which attempt
to represent all details because of their simple reconstruction loss.
Additionally, \gls{cpc} models do not need to reconstruct the input signal like
autoencoder methods. This results in improved computational efficiency which is
significant for high-dimensional signals such as raw waveforms.

The resulting \gls{vqcpc} features are input to a predicting function. Two
predicting functions are considered in this paper to highlight the capacity of
the proposed features to capture useful and accessible information and to show
that our measure is competitive. 

The remainder of this paper is structured as follows.  The computation of the
proposed features and the considered predicting functions are described in
\sref{sec:proposed}. The experiments\footnote{Tools to generate the datasets and
reproduce the experiments are made available online:
\url{https://github.com/vvvm23/stoi-vqcpc}}, including the used datasets of
simulated data and considered benchmark, are described in
\sref{sec:experiments}. The results are presented in \sref{sec:results} and
\sref{sec:conclusion} concludes the paper.

\section{Proposed Method}
\label{sec:proposed}
The proposed method aims at estimating the speech intelligibility from an $\nch$
channel\footnote{Our method can be used on an arbitrary number of channels, but
we simply use $\nch=2$ throughout.}  signal $\timemic{\chind}{\sampind}$, of
length $\nSamples$ and sampling frequency $\fs$, where $\chind$ and $\sampind$
denote the channel index and sample index respectively.  Such a signal can be
modelled as:
\begin{equation}
	\timemic{\chind}{\sampind} = \timea{\sampind} \Conv \timerir{\chind}{\sampind} + \timen{\chind}{\sampind},
	\label{eq:signalmodel}
\end{equation}
where $\timea{\sampind}$ denotes the anechoic speech signal,
$\timerir{\chind}{\sampind}$ denotes the \gls{rir} between the speech source and
the microphone and $\timen{\chind}{\sampind}$ denotes an additive noise signal.
Aiming at non-intrusive prediction, the proposed method estimates the speech
intelligibility from $\timemic{\chind}{\sampind}$ only without knowledge of
$\timea{\sampind}$.  This prediction relies on first computing a sequence of
features to be input to the predicting function. 

\subsection{Feature computation}

The microphone signal is divided into $\nFrames=\ceil{\nSamples/ \hop}$
overlapping frames of length $\frameLength$, where $\hop$ denotes the hop
length. The samples in each $\frameind$\textsuperscript{th} frame are used to construct a vector
of length $\nch \cdot \frameLength$:
\begin{align}
		\micvec{\frameind}=
	\big[
	\timemic{0}{\frameind \hop},\timemic{0}{\frameind \hop + 1}, &\ \ldots \ ,\timemic{0}{\frameind \hop + \frameLength -1},\notag\\
	&\ \ldots\\
	\timemic{\nch-1}{\frameind \hop},\timemic{\nch-1}{\frameind \hop + 1},&\ \ldots\ ,\timemic{\nch-1}{\frameind \hop + \frameLength -1}
					\Transpose{\big]},\notag
\end{align}
resulting in the time-ordered sequence of $\nFrames$ vectors:
\begin{equation}
 \micseq = \big\{ \micvec{0}, \ \micvec{1}, \ \ldots\ ,\ \micvec{\nFrames-1}  \big\}.
\end{equation}
The feature computation results in the sequence:
\begin{equation}
 \featseq = \big\{ \featvec{0}, \ \featvec{1}, \ \ldots\ ,\ \featvec{\nFrames-1}  \big\},
 \label{eq:featuresequence}
\end{equation}
where each vector of length $\nFeats$ is defined as:
\begin{equation}
	\featvec{\frameind} = \big[
	\featcoef{\frameind}{0}, \featcoef{\frameind}{1}, \ \ldots \ , \featcoef{\frameind}{\nFeats - 1}
	\Transpose{\big]},
\end{equation}
where $\featcoef{\frameind}{\featind}$ denotes the $\featind^\textrm{th}$ feature
coefficient extracted from the $\frameind^\textrm{th}$ frame. The feature extraction is
typically designed such that $\nFeats < \nch \cdot \frameLength$
and learns to extract sequences $\featseq$ that
maximise the mutual information between the input and output sequences: 
\begin{equation}
	\mutualInfo{\micseq}{\featseq} = \Sum{\micseq, \featseq}{}{
	\jointProba{\micseq}{\featseq}
	\logTen{
	\frac{\conditionalProba{\micseq}{\featseq}}{\proba{\micseq}}
	}
	}.
	\label{eq:mi}
\end{equation}

To do so, \gls{vq} and \gls{cpc} methods are used to compute the sequence $\featseq$ as a
latent representation of the input sequence $\micseq$.
The computation of these \gls{vqcpc} features consists of three main components:
a non-linear encoder, a \gls{vq} codebook, and an autoregressive aggregator.   

First, the non-linear encoder $\encoder{\cdot}$ maps $\micseq$ to an
intermediate latent representation $\latentseq$:
\begin{equation}
	\encoder{\micseq} = \latentseq =  \big\{ \latentvec{0}, \
    \latentvec{1}, \ \ldots\ ,\ \latentvec{\nFrames-1}  \big\},
\end{equation}
where $\latentvec{\latentind}$ denotes the $\latentind^{\textrm{th}}$
vector, each of length $\embeddinglength$. \gls{vq} is applied to map each
vector in $\latentseq$ to an embedding vector from a finite codebook $\codebook$
yielding the sequence:
\begin{equation}
 \vqseq = \big\{ \vqvec{0}, \ \vqvec{1}, \ \ldots\ ,\ \vqvec{\nFrames-1}  \big\}, 
\end{equation}
where each $\latentind^{\textrm{th}}$ vector $\vqvec{\latentind}$ is computed as:

\begin{equation}\label{eq:codebook}
\vqvec{\latentind} = \quantizer{\latentvec{\latentind}} = \argmin{
|| \latentvec{\latentind} - \codebookvec{\tmpind} ||_{2}
}{
\codebookvec{\tmpind} \in \codebook
}
\end{equation}
Where $\codebookvec{\tmpind}$ denotes the $\tmpind^\textrm{th}$ in the $\nCodebook$
embedding vectors of the codebook.  Finally, an autoregressive aggregator
$\aggregator{\cdot}$ is applied to compute each vector from the sequence in
\eref{eq:featuresequence} as:
\begin{equation}
	\featvec{\frameind} = \aggregator{\vqvec{\latentind \leq \frameind}}.
\end{equation}

\subsection{VQ-CPC training} 
Training of $\encoder{\cdot}$, $\quantizer{\cdot}$ and $\aggregator{\cdot}$ is
conducted end-to-end to maximize the mutual information defined in \eref{eq:mi}.
The proposed approach follows the method in~\cite{oord2019} with
additional loss terms to support the added \gls{vq} codebook~\cite{oord2018}.
To encourage shared information to be encoded, each vector $\featvec{\frameind}$
is used to predict $\vqvec{\frameind+\stepind}$ for up to $\stepind$ steps in
the future.
However, rather than modelling the distribution $\conditionalProba{\micvec{\frameind + \stepind}}{\featvec{\frameind}}$,
the proposed method models the density ratio defined as:
\begin{equation}\label{eq:ratio}
    \densityRatio{\stepind}{\micvec{\frameind + \stepind}}{\featvec{\frameind}}
    \propto
    \frac{
    \conditionalProba{\micvec{\frameind + \stepind}}{\featvec{\frameind}}
    }{
    \proba{\micvec{\frameind + \stepind}}
    }.
\end{equation}
The density ratio $ \densityRatio{\stepind}{\micvec{\frameind +
\stepind}}{\featvec{\frameind}}$ may be unnormalized and, in this paper, is
computed as:
\begin{equation}\label{eq:score}
    \densityRatio{\stepind}{\micvec{\frameind + \stepind}}{\featvec{\frameind}} =
    \Exp{\Transpose{\vqvec{\frameind + \stepind}} \projection{\stepind} \featvec{\frameind}  },
\end{equation}
where $\projection{\stepind}$ denotes a learned linear projection and
$\vqvec{\frameind + \stepind}$ is the output of the encoder corresponding to
$\micvec{\frameind + \stepind}$, used as a proxy for more efficient computation of
the ratio.
Using this definition, the encoder and aggregator are trained by minimising the
InfoNCE loss $\mathcal{L}$, based on noise-contrastive estimation and importance
sampling:
\begin{equation}\label{eq:loss}
    \mathcal{L} = \beta \cdot \mathcal{L}_\textrm{vq} + \frac{1}{k} \sum_{i=1}^k \mathcal{L}_i,
\end{equation}
where $\mathcal{L}_\textrm{vq}$ denotes the weighted \gls{vq} commitment loss
defined as:
\begin{equation}\label{eq:vq}
    \mathcal{L}_\textrm{vq} = \frac{1}{\nFrames}\sum\limits_{\ell = 0}^{\nFrames - 1}
    ||\latentvec{\latentind} - \textrm{sg}[\codebookvec{\tmpind}]||_2^2
\end{equation}
where $\codebookvec{\tmpind}$ is the corresponding embedding vector of
$\latentvec{\latentind}$ and
$\textrm{sg}[\cdot]$ is the stop-gradient operator \cite{oord2018} and:
\begin{equation}\label{eq:nce}
    \mathcal{L}_k = - \mathop\mathbb{E}\limits_X \left[ \log
        \frac{\sigma_k(\micvec{\frameind + \stepind},
    \featvec{\frameind})}{\sum_{\micvec{j} \in X} \sigma_k(\micvec{j},
\featvec{\frameind})} \right],
\end{equation}
where $X$ is a set of many negative samples drawn from $p(\micvec{\frameind +
\stepind})$ and one positive
sample drawn from $p(\micvec{\frameind + \stepind} \vert \featvec{\frameind})$ \cite{oord2019}.  The codebook
embedding vectors are updated using \gls{ema} as described in \cite{oord2018}.

\subsection{Intelligibility score predictor}

The computation of the \gls{vqcpc} features does not rely on any assumptions
about the downstream task for which these features are used.  In this paper, the
sequence $\featseq$ is input to a predicting function for the purpose of
\gls{si} prediction. Two different predicting functions are considered.

The first considered predicting function uses each vector $\featvec{\frameind}$
as input to a single shared linear layer in order to compute a per-frame score. The
score assigned to the complete sequence is the mean of the scores computed from
each vector. This simple predicting function is used to demonstrate how
easily accessible information about \gls{si} is when using the \gls{vqcpc}
features. This predicting function is referred to as \bsq{Small} in the
remainder of the paper.

The second considered predicting function first builds a global representation
using \gls{seqpool} methods originally used for the classification
of images~\cite{hassani2021}.  A global representation is computed by applying
\gls{seqpool} to each vector in the sequence $\featseq$.  In this case
\gls{seqpool} inputs each vector $\featvec{\frameind}$ to a linear layer that
outputs a scalar before applying softmax to the computed scalars, forming
weightings for each frame.  The weighted sum of each vector is then computed, 
forming the global representation.  This global representation is then input to
a small \gls{mlp} to compute the estimated speech intelligibility score assigned
to the sequence $\featseq$. This predicting function is referred to as
\bsq{Pool} in the remainder of the paper.

Both Small and Pool are trained to minimise the \gls{mse} between the estimated
and true speech intelligibility scores (see \sref{sec:experiments}).

\section{Experimental Setup}
\label{sec:experiments}
\subsection{Generated datasets}

For training of both the \gls{vqcpc} model and the predicting functions,
training and development datasets of binaural signals are generated.  All
signals have a sampling frequency $\fs=$\, 16\,kHz and are generated as per
\eref{eq:signalmodel}. The clean anechoic speech is extracted from either the
360 hour training set or the 5 hour development set from the LibriSpeech
corpus~\cite{panayotov2015}.  Reverberant speech is generated by convolving each
utterance of clean speech with a \gls{brir} randomly selected from the Aachen
Impulse Response Database~\cite{jeub2009}.  For each reverberant utterance, two
different noise segments of the same length are selected from the noise signals
in the MUSAN database~\cite{snyder2015}.  These two signals are used to generate
the two-channel noise signal of a spherically isotropic noise field using the
method from~\cite{habets2008}.  Finally, this generated noise signal is added to
the reverberant signal after being scaled to a chosen \gls{snr}, randomly
selected between -10\,dB and 30\,dB, measured in the first channel according
to~\cite{ITU_T_P56}. In the training and development sets, this process is
repeated three times for each clean speech utterance. 

Additionally, two testing datasets are generated, hereafter denoted \bsq{\testA}
and \bsq{\testB}.  The signals in \testA{} are generated using the same method, as well as noise and \gls{brir} datasets,
as for the training and development sets but using clean speech from the test
split of the LibriSpeech corpus. The signals in \testB{} are generated by
convolving the speech signals used as target utterances in the first Clarity
Challenge~\cite{graetzer2021,graetzer2022} with \glspl{brir} randomly selected from the \glspl{brir} available in~\cite{kayser2009} recorded in either a cafeteria or a
courtyard.  In this case, two-channel noise signals recorded at the same
location are used and added to the reverberant signals with an \gls{snr}
randomly selected and measured.

A total of 1090.8, 16.2, 5.4 and 10.4 hours of data are generated in the training,
development, \testA{} and \testB{} dataset, respectively.  
Labelling this large amount of data in terms of intelligibility would be a daunting task and the experiments aim mostly at evaluating the use of the proposed features.
Consequently, we labeled all signals with an intrusive measure known to highly correlate with intelligibility
and the ground truth is here defined as the \gls{dbstoi} computed using the clean
reverberant signal and the noisy reverberant signal as
input~\cite{andersen2018a}.

\begin{figure*}[!th]
    \centering
	\hspace{-0.15cm}\input{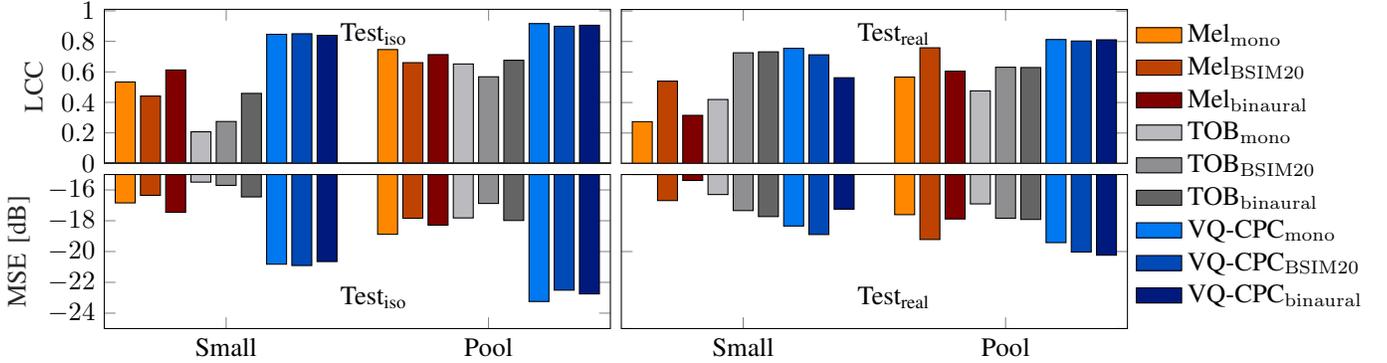}
    \caption{Performance of the proposed \acrshort{vqcpc} and considered benchmark features on the \testA{} (left) and \testB{} (right) datasets when using either the Small or Pool predicting function.
    All correlations appeared significant, with p-values inferior to 0.01.}
    \label{fig:results}
\end{figure*}

\subsection{Parameters of proposed method}

For training, we use $\micseq$ of length $T=40960$ as input to the encoder
$\encoder{\cdot}$.  The encoder has a frame length and hop size of 25~ms and
10~ms respectively. It is implemented as a series of five convolutional blocks,
each consisting of a one-dimensional convolutional layer with 256 filters, a
dropout layer~\cite{srivastava2014}, batch normalisation~\cite{ioffe2015} and
the \gls{relu} activation function.  The strides for each block are $[5,4,2,2,2
]$ and the kernel sizes are $[10,8,4,4,4]$.  \gls{vq} is applied using a
codebook of 512 vectors of dimensionality 128, with the commitment loss defined
as in~\eqref{eq:vq}.
The aggregator $\aggregator{\cdot}$ is implemented as a two-layer \gls{gru}
\cite{chung2014} with 128 hidden channels. Hence, in our experiments, $\nFeats =
\embeddinglength$.
The InfoNCE loss is computed using $10$ negative samples and $\stepind=$~12
steps.
Augmentation is applied as random channel and polarity swapping, additive noise
and random audio gain.
 All resulting sequences $\featseq$ (with 
$\nFeats=$\,128) in the training set are used to train the considered
predicting functions. \gls{vqcpc} features are extracted from \testA{} and
\testB{} using the complete \gls{vqcpc} model trained on the training set.

The Small predicting function consists of a single layer mapping each feature
vector of length $\nFeats$ to a single element followed by the sigmoid
activation function. The Pool predictor is implemented as a single shared
linear layer to compute the weighting and a \gls{mlp} with one hidden layer
 of
size 2$\nFeats$. The hidden layer uses \gls{relu} as its non-linearity and the
output layer consists of a single element followed by the sigmoid activation
function.

Training and testing of the \gls{vqcpc} model and predicting functions were
implemented in PyTorch~\cite{paszke2019}. The total number of trainable network
weights in the \gls{vqcpc} model is $\approx 1.74 \times 10^6$.

\subsection{Benchmark and figures of merit}

The performance of the proposed \gls{vqcpc} is measured in terms of \gls{lcc} and \gls{mse} between the ground truth and the output of the
predicting functions. The experiments aim to quantify the ability of \gls{vqcpc}
features to represent information useful for speech intelligibility prediction.
To this end, their performance is compared with the use of mel-spectrogram (Mel), with deltas and double-deltas, 
and with envelopes extracted in third-octave bands (TOB) similarly as used in~\cite{andersen2018b}.
All features are extracted from either the first channel (mono), concatenated from both channels (binaural) or extracted from the single-channel signal computed using a blind binaural preprocessing
stage~\cite{hauth2020} (BSIM20). The type of signal is indicated in subscript in the following.
All features are computed using the same frame length and hop size as the \gls{vqcpc}.
For all considered features, the predicting
functions Small and Pool are trained using the same dataset as for \gls{vqcpc}.

\section{Results}
\label{sec:results}


All results are depicted in \fref{fig:results}.
On \testA{}, \gls{vqcpc} features yield the best performance regardless of the type of signals from which they are computed, when using either the Small or the Pool predicting function.
Using Small and \gls{vqcpc} features
yields a \gls{lcc} 0.84 and an \gls{mse}
of -20.7\,dB.  Using Pool and \gls{vqcpc} features yields a \gls{lcc} of 0.94 and an \gls{mse} of -22.7\,dB.  In
contrast with the other considered features, the difference in performance
between the use of Small and Pool is rather modest. Success applying Small
suggests that the \gls{vqcpc} features contain easily accessible information
about the intelligibility of speech.

On \testB{}, the performance of all combinations of features and predicting
functions decreases, as expected. It can however be noted that the TOB features,
that performed the least satisfactorily on the less challenging \testA{}, outperform Mel on \testB{}. This seems to confirm their suitability in realistic scenarios~\cite{andersen2018b}.
The proposed \gls{vqcpc} features remain the best performing of the considered features.
Using
\gls{vqcpc} features computed from binaural signals as input to the Pool predicting function yields
\gls{lcc} of 0.81 and an
\gls{mse} of -20.2\,dB.


A more
powerful predictor such as STOI-Net~\cite{zezario2020} could be
improve performance further, but we emphasize the purpose of our study is to show that good
performance can be obtained with \gls{vqcpc} features alone. Though the
\gls{vqcpc} were here proposed to predict intelligibility from binaural signals,
the difference in performance between VQ-CPC$_\mathrm{mono}$,
VQ-CPC$_\mathrm{BSIM20}$ and VQ-CPC$_\mathrm{binaural}$ is modest. Further
experimentation, e.g., using intelligibility scores as ground truth rather than
an intrusive measures, are needed to determine if the VQ-CPC features do capture
information such as binaural cues. Regardless, the difference in network size
between the various \gls{vqcpc} models is negligible.

\section{Conclusion}
\label{sec:conclusion}
This paper proposes to use \gls{vqcpc} features as input to a trained
neural network to non-intrusively predict intelligibility from binaural signals.
The performance of the proposed measure is assessed in terms of correlation and \gls{mse}. Results show that the \gls{vqcpc} features are
effective in encoding readily accessible information relevant for \gls{si}
prediction and the features outperform all considered
benchmarks. This is despite \gls{vqcpc} features not relying on any assumptions
about the downstream task of \gls{si} prediction.

\balance
\bibliographystyle{utilities/IEEEbib}
\bibliography{utilities/refs}

\begin{thebibliography}{10}

\bibitem{doire2017b}
C.~S.~J. Doire, M.~Brookes, and P.~A. Naylor,
\newblock ``Robust and efficient {B}ayesian adaptive psychometric function
  estimation,''
\newblock {\em J. Acoust. Soc. Am.}, vol. 141, no. 4, pp. 2501--2512, 2017.

\bibitem{bronkhorst2000}
A.~W. Bronkhorst,
\newblock ``The cocktail party phenomenon: A review of research on speech
  intelligibility in multiple-talker conditions,''
\newblock {\em Acta Acustica united with Acustica}, vol. 86, no. 1, pp.
  117--128, Jan. 2000.

\bibitem{French1947}
N.~French and J.~C. Steinberg,
\newblock ``Factors governing the intelligibility of speech sounds,''
\newblock {\em J. Acoust. Soc. Am.}, vol. 19, no. 1, pp. 90--119, Nov. 1947.

\bibitem{steeneken1980}
H.~J.~M. Steeneken and T.~Houtgast,
\newblock ``A physical method for measuring speech-transmission quality,''
\newblock {\em J. Acoust. Soc. Am.}, vol. 67, no. 1, pp. 318--326, Jan. 1980.

\bibitem{ANSI_S3_5_1997}
ANSI,
\newblock ``Methods for the calculation of the speech intelligibility index,''
\newblock ANSI Standard S3.5--1997 (R2007), American National Standards
  Institute, 1997.

\bibitem{taal2011}
C.~H. Taal, R.~C. Hendriks, R.~Heusdens, and J.~Jensen,
\newblock ``An algorithm for intelligibility prediction of time-frequency
  weighted noisy speech,''
\newblock {\em {IEEE} Trans. Audio, Speech, Lang. Process.}, vol. 19, no. 7,
  pp. 2125--2136, Sept. 2011.

\bibitem{taghia2014}
J.~Taghia and R.~Martin,
\newblock ``Objective intelligibility measures based on mutual information for
  speech subjected to speech enhancement processing,''
\newblock {\em {IEEE} Trans. Audio, Speech, Lang. Process.}, vol. 22, no. 1,
  pp. 6--16, Jan. 2014.

\bibitem{andersen2017}
A.~H. Andersen, J.~M. de~Haan, Z.-H. Tan, and J.~Jensen,
\newblock ``A non-intrusive short-time objective intelligibility measure,''
\newblock in {\em Proc. {IEEE} Intl. Conf. on Acoustics, Speech and Signal
  Processing ({ICASSP})}, New Orleans, USA, Mar. 2017, pp. 5085--5089.

\bibitem{sorensen2018}
C.~S{\o}rensen, M.~S. Kavalekalam, A.~Xenaki, J.~B. Boldt, and M.~G.
  Christensen,
\newblock ``Non-intrusive codebook-based intelligibility prediction,''
\newblock {\em Speech Communication}, vol. 101, pp. 85--93, 2018.

\bibitem{spille2018}
C.~Spille, S.~D. Ewert, B.~Kollmeier, and B.~T. Meyer,
\newblock ``Predicting speech intelligibility with deep neural networks,''
\newblock {\em Computer Speech and Language}, vol. 48, pp. 51--66, 2018.

\bibitem{schaedler2020}
R.~Sch\"{a}dler, M.\, D.~H\"{u}lsmeier, A.~Warzybok, and B.~Kollmeier,
\newblock ``Individual aided speech-recognition performance and predictions of
  benefit for listeners with impaired hearing employing {FADE},''
\newblock {\em Trends in Hearing}, vol. 24, 2020.

\bibitem{zezario2020}
R.~E. Zezario, S.-W. Fu, C.-S. Fuh, Y.~Tsao, and H.-M. Wang,
\newblock ``{STOI}-{N}et: A deep learning based non-intrusive speech
  intelligibility assessment model,'' 2020,
\newblock arXiv:2011.04292.

\bibitem{durlach1963}
N.~I. Durlach,
\newblock ``Equalization and cancellation theory of binaural masking-level
  differences,''
\newblock {\em J. Acoust. Soc. Am.}, vol. 35, no. 8, pp. 1206--1218, 1963.

\bibitem{beutelmann2010}
R.~Beutelmann, T.~Brand, and B.~Kollmeier,
\newblock ``Revision, extension, and evaluation of a binaural speech
  intelligibility model,''
\newblock vol. 127, no. 4, pp. 2479--2497, 2010.

\bibitem{lavandier2010}
M.~Lavandier and J.~F. Culling,
\newblock ``Prediction of binaural speech intelligibility against noise in
  rooms,''
\newblock vol. 127, no. 1, pp. 387--399, 2010.

\bibitem{andersen2018a}
A.~H. Andersen, J.~M. de~Haan, Z.-H. Tan, and J.~Jensen,
\newblock ``Predicting the intelligibility of noisy and non-linearly processed
  binaural speech,''
\newblock {\em {IEEE} Trans. Audio, Speech, Lang. Process.}, vol. 26, no. 10,
  pp. 1925--1939, Oct. 2018.

\bibitem{rossbach2021}
J.~Ro{\ss}bach, S.~R\"{o}ttges, C.~F. Hauth, T.~Brand, and B.~T. Meyer,
\newblock ``Non-intrusive binaural prediction of speech intelligibility based
  on phoneme classification,''
\newblock in {\em Proc. {IEEE} Intl. Conf. on Acoustics, Speech and Signal
  Processing ({ICASSP})}, Toronto, Ontario, Canada, June 2021, pp. 396--400.

\bibitem{hauth2020}
C.~F. Hauth, S.~C. Berning, B.~Kollmeier, and T.~Brand,
\newblock ``Modeling binaural unmasking of speech using a blind binaural
  processing stage,''
\newblock {\em Trends in Hearing}, vol. 24, Jan. 2020.

\bibitem{hermansky2013}
H.~Hermansky, E.~Variani, and V.~Peddinti,
\newblock ``Mean temporal distance: Predicting {ASR} error from temporal
  properties of speech signal,''
\newblock in {\em Proc. {IEEE} Intl. Conf. on Acoustics, Speech and Signal
  Processing ({ICASSP})}, Vancouver, Canada, May 2013, pp. 7423--7426.

\bibitem{sharma2016}
D.~Sharma, Y.~Wang, P.~A. Naylor, and M.~Brookes,
\newblock ``A data-driven non-intrusive measure of speech quality and
  intelligibility,''
\newblock {\em Speech Communication}, vol. 80, pp. 84--94, 2016.

\bibitem{andersen2018b}
A.~H. Andersen, J.~M. de~Haan, Z.-H. Tan, and J.~Jensen,
\newblock ``Nonintrusive speech intelligibility prediction using convolutional
  neural networks,''
\newblock {\em {IEEE} Trans. Audio, Speech, Lang. Process.}, vol. 24, no. 11,
  pp. 1908--1920, July 2018.

\bibitem{oord2019}
A.~van~den Oord, Y.~Li, and O.~Vinyals,
\newblock ``Representation learning with contrastive predictive coding,'' 2019,
\newblock arXiv:1807.03748.

\bibitem{vanniekerk2020}
B.~van Niekerk, L.~Nortje, and H.~Kamper,
\newblock ``Vector-quantized neural networks for acoustic unit discovery in the
  {Z}ero{S}peech 2020 challenge,'' 2020,
\newblock arXiv:2005.09409.

\bibitem{schneider2019}
S.~Schneider, A.~Baevski, R.~Collobert, and M.~Auli,
\newblock ``{w}av2vec: Unsupervised pre-training for speech recognition,''
  2019,
\newblock arXiv:1904.05862.

\bibitem{baevski2020}
A.~Baevski, S.~Schneider, and M.~Auli,
\newblock ``{v}q-wav2vec: Self-supervised learning of discrete speech
  representations,'' 2020,
\newblock arXiv:1910.05453.

\bibitem{oord2018}
V.~van~den Oord, O.~Vinyals, and K.~Kavukcuoglu,
\newblock ``Neural discrete representation learning,'' 2018,
\newblock arXiv:1711.00937.

\bibitem{hassani2021}
A.~Hassani, S.~Walton, N.~Shah, A.~Abuduweili, J.~Li, and H.~Shi,
\newblock ``Escaping the big data paradigm with compact transformers,'' 2021,
\newblock arXiv:2104.05704.

\bibitem{panayotov2015}
V.~Panayotov, G.~Chen, D.~Povey, and S.~Khudanpur,
\newblock ``{L}ibri{S}peech: An {ASR} corpus based on public domain audio
  books,''
\newblock in {\em Proc. {IEEE} Intl. Conf. on Acoustics, Speech and Signal
  Processing ({ICASSP})}, South Brisbane, Queensland, Australia, Apr. 2015, pp.
  5206--5210.

\bibitem{jeub2009}
M.~Jeub, M.~Schaefer, and P.~Vary,
\newblock ``A binaural room impulse response database for the evaluation of
  dereverberation algorithms,''
\newblock in {\em Proc. {IEEE} Intl. Conf. Digital Signal Processing ({DSP})},
  Santorini, Greece, July 2009.

\bibitem{snyder2015}
D.~Snyder, G.~Chen, and D.~Povey,
\newblock ``{MUSAN}: a music, speech, and noise corpus,'' 2015,
\newblock arXiv:1510.08484v1.

\bibitem{habets2008}
E.~A.~P. Habets, I.~Cohen, and S.~Gannot,
\newblock ``Generating nonstationary multisensor signals under a spatial
  coherence constraint,''
\newblock {\em J. Acoust. Soc. Am.}, vol. 124, no. 5, pp. 2911--2917, Nov.
  2008.

\bibitem{ITU_T_P56}
ITU-T,
\newblock ``Objective measurement of active speech level,'' Mar. 1993.

\bibitem{graetzer2021}
S.~Graetzer, J.~Barker, T.~J. Cox, M.~Akeroyd, J.~F. Culling, G.~Naylor,
  E.~Porter, and R.~Viveros-Mu\~{n}oz,
\newblock ``{C}larity-2021 challenges: {M}achine learning challenges for
  advancing hearing aid processing,''
\newblock in {\em Proc. Interspeech}, Brno, Czech Republic, Aug. 2021.

\bibitem{graetzer2022}
S.~Graetzer, M.~Akeroyd, J.~Barker, T.~J. Cox, J.~F. Culling, G.~Naylor,
  E.~Porter, and R.~Viveros-Mu\~{n}oz,
\newblock ``Dataset of {B}ritish {E}nglish speech recordings for
  psychoacoustics and speech processing research: {T}he clarity speech
  corpus,''
\newblock {\em {D}ata in {B}rief}, vol. 41, Apr. 2022.

\bibitem{kayser2009}
H.~Kayser, S.~D. Ewert, J.~Anem\"{u}ller, T.~Rohdenburg, V.~Hohmann, and
  B.~Kollmeier,
\newblock ``Database of multichannel in-ear and behind-the-ear head-related and
  binaural room impulse responses,''
\newblock {\em {EURASIP} Journal on Advances in Signal Processing}, vol. 2009,
  pp. 1--10, July 2009.

\bibitem{srivastava2014}
N.~Srivastava, G.~Hinton, A.~Krizhevsky, I.~Sutskever, and R.~Salakhutdinov,
\newblock ``Dropout: A simple way to prevent neural networks from
  overfitting,''
\newblock {\em Journal of Machine Learning Research}, vol. 15, pp. 1929--1958,
  2014.

\bibitem{ioffe2015}
S.~Ioffe and C.~Szegedy,
\newblock ``Batch normalization: Accelerating deep network training by reducing
  internal covariate shift,'' 2015,
\newblock arXiv:1502.03167.

\bibitem{chung2014}
J.~Chung, C.~Gulcehre, K.~Cho, and Y.~Bengio,
\newblock ``Empirical evaluation of gated recurrent neural networks on sequence
  modeling,'' 2014,
\newblock arXiv:1412.3555.

\bibitem{paszke2019}
A.~Paszke, S.~Gross, F.~Massa, A.~Lerer, J.~Bradbury, G.~Chanan, T.~Killeen,
  Z.~Lin, N.~Gimelshein, L.~Antiga, A.~Desmaison, A.~K\"{o}pf, E.~Yang,
  Z.~De{V}ito, M.~Raison, A.~Tejani, S.~Chilamkurthy, B.~Steiner, L.~Fang,
  J.~Bai, and S.~Chintala,
\newblock ``{P}y{T}orch: An imperative style, high-performance deep learning
  library,'' 2019,
\newblock arXiv:1912.01703.

\end{thebibliography}

\end{document}